\documentclass[11pt]{article}
\usepackage[utf8]{inputenc}
\usepackage[T1]{fontenc}
\usepackage[english]{babel}
\usepackage{amsmath,amsthm,amssymb}
\usepackage{mathtext}
\usepackage{mathptmx}
\usepackage{multirow}
\usepackage{longtable}
\usepackage{geometry}
\usepackage{natbib}
\usepackage[colorlinks=true,allcolors=blue]{hyperref}
\usepackage[affil-it]{authblk}

\title{Research Methods of Assessing Global Value Chains}
\author{Sourish Dutta \\ {\small{PhD Student}}}
\affil{Centre for Development Studies \\ Trivandrum, Kerala}
\date{}

\begin{document}

\maketitle

\section{Quantification of GVC Links}

This section would use several statistical measures to examine the degree of India's GVC engagement. Those measures can be classified according to the levels of vertical integration, such as measures of backward integration, measures of forward integration, and some other statistical measures. I have summarised the main measures of forward and backward dimensions as well as the other crucial measures of India's GVC engagement in the following sections (data for all measures are easily available for India).

\subsection{GVC Participation Using Gross Trade Data}

Investigating India's potential in GVCs, at first, we have to use the gross trade flows. It requires to looking at the top export and import products, classified at the most disaggregated level. Execution of such initial insights increases the relevance of GVC analysis. In this first-cut analysis, we will handle three issues: 

\begin{enumerate}
              \item Compare product-level imports with export values, volumes, and prices of the top traded products (whether exports and imports follow different distributions and the values or volumes traded have distinct growth or level) to examine the degree of transformation within domestic segment of India's major GVCs \citep{10}.
              
              \item Use informed classifications (based on the WITS value chain category, OECD Bilateral Trade Database by industry and end-use category, and UN Classification of Broad Economic Categories) to extract as much information as possible from gross trade data \citep{sturgeon2011mapping,athukorala2011production}. Regrouping data in meaningful clusters or categorised by informed classifications focusing on consumer goods, capital goods, intermediates, and raw materials with sector analysis is beneficial \citep{10}. 
              
              \item Document trade flows at the sub-national level to account India's degree of transformation within border when available.
              
\end{enumerate}

These measures do not reveal whether the inputs are used domestically or exported. The following measures address this topic, focusing on imported inputs or foreign value added embodied in gross exports.

\subsection{Buyer's Perspective: Methods of Backward Links}

This section onward, different measures suggest ways identify the extent to which India and its peers source -- domestically or internationally -- the intermediates they use in exporting, which will provide the first indication of their participation in GVCs. The section then shows ways to quantify the domestic value added embodied in countries’ exports \citep{48}. Key questions for sourcing dimension are: Where are India's exports made, and where is the value created? 
\begin{itemize}
    \item \underline{\textbf{Imported Inputs Embodied in Gross Exports:}}
\begin{enumerate}
\item \textbf{I2E in Intermediate or Total Imports} 
      
 Measuring the India's intermediate imports embodied in its gross exports as a percentage of the India's total intermediate imports. Major businesses in GVCs perceive imports of goods and services as being important or critical for their exports \citep{oecd-wto}.
     
\item \textbf{I2E by Source Country} 
      
 A very useful indicator of GVC participation is the origin of the imported inputs embodied in India's gross exports.
      
\item \textbf{Distinguishing between Domestic and Foreign Value
Added in Imports} 

Imported inputs may contain domestic value added that is exported to a foreign location, processed, and re-imported. Re-importing and re-exporting can be quite important for some industries in India.
\end{enumerate}
  
  \item \underline{\textbf{Value Added in Gross Exports:}}
  
\begin{enumerate}
  
 \item \textbf{Decomposition of Domestic Value Added}

The first-pass indicator simply distinguishes between domestic and foreign value added, usually expressed as a share of gross exports. The second pass digs deeper into where the domestic value added is actually created. This method breaks down the total domestic value added into three parts i.e. domestic value added in the particular sector --  Direct domestic value added, domestic value added in upstream sectors supplying the sector with parts -- Indirect domestic value added, and domestic value added in intermediates first shipped abroad for further processing and then re-imported --  Re-imported intermediates

\item \textbf{Full Decomposition of Foreign Value Added}

The breakdown of foreign value added into source countries or industries is useful from a buyer's perspective because it identifies which foreign sources add the most value to its exports.
      
  \end{enumerate}
   
\item \underline{\textbf{Length of Sourcing Chains:}}
    
This measure that reflects such multi-country considerations captures by looking at a recursive measure of I2E on the sourcing side is the length of value chain sourcing \citep{fally2011fragmentation, de2013mapping}. The TiVA data provide a handy means for comparing the average number of production stages in a given industry and country. An increase in GVC length over time suggests that the value chain has become more complex, with stages done in more countries.
\end{itemize}

\underline{\textbf{Data Sources:}} Gross import data (UN Comtrade, BACI, WITS), categorized using informed classifications (BEC, parts and components, technical classifications); International I-O data (WIOD, TiVA, World Bank Export Value Added database); Enterprise surveys or other firm-level surveys.

\subsection{Seller's Perspective: Methods of Forward Links}

Key questions for the selling dimension are: Who are the ultimate customers for India's value added, and to what countries is India exporting its value added? India, for example, exports iron ore to China, but part of that product ends up in the United States and Germany rather than China. That is the seller's perspective.

\begin{itemize}
    
\item \underline{\textbf{Intermediates in Output or Gross Exports:}}
    
A first basic measure of India's involvement in the production of inputs, as opposed to final goods, is the share of intermediates in gross output. The share of intermediates in gross exports takes the exporting perspective into account. The measure for India and Indian sectors, and relative to peers can provide a first-pass indication of whether India has become a more important supplier in GVCs.
    
\item \underline{\textbf{I2E Trade in Gross Exports:}}
    
The indicator importing to exports (I2E) in gross exports measures intermediates sold by a country to a buyer for use in the buyer's exports (I2E from the buyer’s perspective) as a percentage of the seller's gross exports.

\item \underline{\textbf{Domestic Value Added in Gross Exports of
Third Countries:}}

It indicates the contribution of domestically produced intermediates to exports in third countries. The only difference is that this indicator accounts only for the seller's intermediates that are domestically produced, whereas in the previous case the intermediates could also contain some foreign value added \citep{gaulier2013wake}.

\item \underline{\textbf{Value Added in Final Domestic Demand:}}

The Trade in Value Added (TiVA) and World Input-Output databases can also provide an understanding of the final consumers of India's value-added activities \citep{oecd}. 
 
\item \underline{\textbf{Length of Selling Chains:}}

GVC length on the sales side measure gauges the "upstreamness" of India's exports roughly, the number of downstream stages between India's producers and final consumers \citep{antras2012measuring, chor2014countries}.

\item \underline{\textbf{Domestic Gap between Buying and Selling Chains}}

A final useful metric is to combine import and export upstreamness to compute the domestic gap between the buying and selling chains of individual sectors.

\end{itemize}

\underline{\textbf{Data Sources:}} Production data (national statistics, UN-Stat manufacturing data set, firm-level data); Gross export data (Comtrade, BACI, WITS), categorized using informed
classifications (broad economic category, parts and components, technical classifications); International I-O data (WIOD, TiVA, World
Bank Export of Value Added database) and National I-O data; Enterprise surveys or other firm-level surveys. 

\subsection{GVC Participation: Methods from Macro to Micro}
\begin{itemize}
\item \underline{\textbf{Additional useful measures of GVC participation:}} 

Beyond the Measures of Buying \& Selling Sides in India's GVCs...

\begin{enumerate}
               \item Illustrating how the buyer and seller dimensions can be combined to quantify an overall indicator i.e. the GVC participation index \citep{koopman2010give}.
               \item Focusing on network metrics. It shows how a country's position overall, in a sector, in a specific GVC, and with respect to individual products can be measured and visualised using network metrics \citep{amador2015age}.
               \item Paying special attention to the role of services in value added \citep{saez2015valuing, francois2015services}.
               \item Measures of direct links in GVCs using firm-level data -- the micro perspective.
    \end{enumerate}
    
\item \underline{\textbf{Links in GVCs Using Firm-Level Measures:}} 

\begin{enumerate}
        \item Multinationals' Share of Inputs from Domestic  Suppliers \citep{blalock2008welfare, smarzynska2004does, havranek2011estimating}\smallskip
        \item Domestic Suppliers' Share of Output to Multinationals\smallskip
        \item Domestic Suppliers' Share of Exports\smallskip
        \item Domestic Producers' Share of Imported Inputs
    \end{enumerate}

\end{itemize}

However, the next phase would follow various econometric tests with customised analysis to India-specific needs and overcoming the limits inherent in the above methodologies.

\section{Determinants of GVC Links}
This section focuses on different determinants of GVC
links at the country, sector, and firm levels. The first step decomposes gross export growth into its components. If gross export growth is accepted as a measure of GVC links on the selling side, the decomposition allows for detecting how much of the value added is generated at home and abroad. The second part adopts two measures of GVC links -- GVC integration at the country or sector level and a GVC participation dummy at the firm level. 

\subsection{Decomposition of Gross Export Growth}

This measure would examine the level of significance of gross export growth onto its components i.e. direct domestic value added embodied in gross exports (intra-sector), indirect (upstream) domestic value added embodied in gross exports, re-imported domestic value added, and foreign value added embodied in gross exports. If gross export growth is accepted as a measure of GVC links on the selling side, the decomposition would allow for detecting how much of the value added is generated across countries (India and its peer countries) or sectors within India. 
           
Now we should adopt two measures of India's GVC links -- GVC integration at the country or sector level and a GVC participation dummy at the firm level. This research, indeed, focuses on different determinants of GVC links at the country, sector, and firm levels.
    
\subsection{Correlations of GVC Integration with Country-Level Characteristics}
Here we focus on the following three country characteristics,
which, according to the economic literature, are important determinants of GVC participation: (1) logistics performance, (2) share of people with a tertiary education in the workforce, and (3) geographical distance to the closest global knowledge centers. However, initial insights can be gathered by assessing the statistical correlation between measures of GVC integration with
those selected indicators at the country level. This analysis
uses the measure of structural integration in GVCs \citep{amador2015age}.

\begin{enumerate}
    \item \textbf{GVC Integration and Logistics Performance}
    
    Good logistics performance is important because key components of GVC production are time sensitive, and reliable connectivity allows firms to connect factories across borders more efficiently. We will use the overall Logistics Performance Index (LPI) to quantify logistics performance. The LPI takes into account a country’s customs efficiency, quality of trade and transport infrastructure, ease of arranging shipments, quality of logistics services, ability to track and trace consignments, and delivery times.
    
    \item \textbf{GVC Integration and Skill Levels}
    
    A skilled workforce is recognised as an important determinant of countries' success in GVCs because it allows producing at the high standards of productivity, efficiency, sophistication, and timeliness required to serve global markets. We will use the share of workers with tertiary education to quantify the skill level.
    
    \item \textbf{GVC Links and Geographical Distance to Knowledge Centers}
    
    Countries closer to the hubs in GVCs and to the global centers of knowledge are favoured by easier access to tacit knowledge. Unlike knowledge embodied in technology, tacit knowledge requires frequent and continued face-to-face interaction between the staff and managers of lead firms or turnkey suppliers and those of other firms in the GVC, and the importance of tacit knowledge increases for more complex tasks. We will use the geographical distance from Germany, Japan, and the United States as a proxy for distance from knowledge centers.
\end{enumerate}

\subsection{Determinants of Firm-Level GVC Entry}
    
Following the literature on the firm-level determinants of exporting, the model includes firm size, firm age, foreign ownership status, as well as measures of workers' skills and productivity as determinants of GVC participation. Here we also try to amend the \cite{roberts1997decision} theoretical model on the determinants of exporting.

\subsection{Determinants of Sector GVC Participation}
If firm-level data -- in particular, information on GVC participation -- are not available, an alternative could be to estimate the impact of the policy determinants on sector GVC participation in India.

\section{Economic Upgrading of GVC Links}
This section focuses on the role of GVC links for economic upgrading. Three main measures of economic upgrading are adopted: (1) growth of domestic value added embodied in gross exports at the sector level in the first section, (2) level of domestic value added at the sector level in the second section, and (3) firm-level labor productivity in the third section. Different measures of GVC links are also explored, including GVC measures of structural integration as buyers and sellers in networks, foreign value added embodied in gross
exports, domestic value added embodied in exports of third countries, GVC participation index, position in GVCs (upstreamness), domestic length of sourcing chains, and share of foreign output in a
sector.

\subsection{Growth of GVC Links and Domestic Value Added in Exports}
Do the intensity and nature of GVC links matter for growth in domestic value added that is exported? This question can be explored through econometric analysis from several angles i.e. (a) whether the degree of structural integration in global value-added trade matters, (b) econometric analysis can be used to investigate
how greater integration of India in GVCs as a buyer -- as opposed to weaker integration as a seller (that is, more unbalanced GVC integration) -- affects domestic value-added growth from gross exports, (c) the analysis can examine more closely the relation between the growth of foreign value added embodied in gross exports and the domestic value-added component, (d) it can look at the role of India's position in the value chain (upstreamness or distance to final demand), and (e) econometrics can be used to investigate the role of the domestic length of the sourcing chains.

\begin{enumerate}
    \item \textbf{Growth of GVC Participation and Domestic Value Added Embodied in Exports}
    \item \textbf{Growth of Balanced GVC Participation and Domestic Value Added Embodied in Gross Exports}
    \item \textbf{Growth of Foreign and Domestic Value Added Embodied in Gross Exports}
    \item \textbf{Growth of Upstreamness and Domestic Value Added Embodied in Gross Exports}
    \item \textbf{Growth of Domestic Length of Sourcing Chains and Domestic Value Added Embodied in Gross Exports}
\end{enumerate}

\subsection{GVC Links and Domestic Value Added}
    
It focuses on the effect of GVC integration, as a buyer and a seller, on domestic value added, also taking into account the mediating role of national policy. Domestic value added is generated by combining labour with capital stock, and is dependent on a country’s technology shifter. The technology shifter is assumed to be a function of international trade and innovation, which is consistent with the trade literature \citep{kummritz2017economic}.
    
\subsection{GVC Participation and Firm-Level Productivity}

\begin{enumerate}
    \item \textbf{Within-Industry Impact of FDI:} 
    
    The method focuses on the within industry impact of foreign output share on domestic firm productivity and the role of mediating factors. Specifically, \cite{farole2014role} ask, what is the potential of global production networks to enhance the productivity of domestic firms?

    \item \textbf{Within-Industry Impact of Structural Integration in GVCs:}
    
    Similarly, the analysis can be used to examine the effect of GVC participation of an industry on a firm’s productivity by merging the \cite{farole2014role} data set with two sector measures of structural integration, computed by \cite{amador2015age}, in India's GVCs.

\end{enumerate}

\section{Social Upgrading of GVC Links}
This section addresses which GVC-oriented industries have a higher demand for labour, such that integrating into GVCs in those sectors
has a greater potential to create jobs and increase household income \citep{18, cali2017much}. We have following measures (using World Bank's LACEX database) that can be used to identify the impact on labour and wages. The measures are categorised in two groups: indirect measures of social upgrading, and direct measures of social upgrading \citep{10}.
    
\subsection{Indirect Measures of Social Upgrading}
This subsection presents indirect measurements of the link between GVC participation and labour market outcomes. The specific sectors that are relevant for participation in GVCs can be analysed using the following methods.

\begin{enumerate}
    \item \textbf{Descriptive Statistics}
    
    Descriptive statistics may be used to assess which sectors are associated with better labour market outcomes. I would examine India and its peer countries' sector averages of the number of employees, wages and salaries, wage rate (wages and salaries divided by the number of employees), or labour share (wages and salaries as a percentage of value added). Such statistics, for example, can be obtained from the United Nations Industrial Development Organisation's Industrial Statistics database.
    
    \item \textbf{Analysis of Employment-Generating Industries and Their Level of GVC Integration}
    
    This analysis may be carried out by running cross-country "controlled correlations" at the sector level, whereby the labour market indicators discussed in the previous section are regressed on indicators of GVC involvement while controlling for other factors, such as region and gross domestic product. The analyst can also run pooled regressions controlling for industry fixed effects to see which industries have more labour-market-enhancing outcomes conditional on GVC involvement.
    
\end{enumerate}

\subsection{Direct Measures of Social Upgrading}
This subsection presents more direct measurements of the link between GVC participation and labour market outcomes by drawing on various indicators already developed in the literature \citep{10}.

\begin{enumerate}
    \item \textbf{Labor Content of Gross Exports}
    
    This direct measure of social upgrading is the labour content of gross exports. The newly developed World Bank data set on LACEX can be used to explore the social upgrading linked to GVC participation \citep{18}. The data set is computed on the basis of the social accounting matrix data available in the Global Trade Analysis Project for intermittent years between 1995 and 2011.
    
    \item \textbf{Labour Component of Domestic Value Added in Exports}
    
    This direct measure of social upgrading, which was developed by the United Nations Conference on Trade and Development \citep{6}, is the labour cost component of domestic value added in exports, which acts as a proxy for the employment generating potential of exports.
    
    \item \textbf{Jobs Sustained by Foreign Final Demand}
    
    This indicator, jobs sustained by foreign final demand, is being developed by OECD-WTO as part of the TiVA database for 40 countries. The indicator calculates the number of jobs in the total economy sustained by foreign final demand, which captures the full upstream impact of final demand in foreign markets on domestic employment. Rather than consider the domestic value added in total exports (as was the basis of the previous indicator), which could be used as intermediates in third countries and be exported as final goods, the indicator considers the domestic value added in foreign final demand.
    
    \item \textbf{Jobs Generated by Foreign Trade in GVCs}

India's participation in GVCs can lead to domestic or foreign labour demand. So this indicator is the number of jobs generated by India's trade in GVCs -- jobs generated domestically and abroad -- using the World Input-Output Database (WIOD). The sources of employment creation from international trade are labour demand from final
goods trade and trade in intermediates, or the result of India's GVC participation \citep{jiang2013capturing}.

\item \textbf{Jobs in GVC Manufacturing}

The jobs in the GVC manufacturing indicator will present a broader
picture of the structure of employment in GVCs within India and its peers using WIOD. It is the most direct measure in the literature of the domestic employment impacts of manufacturing GVC participation The indicator would measure -- directly and indirectly -- the number of GVC jobs involved in the production of final manufacturing goods (also known as manufactures), as well as their sector of employment in India \citep{timmer2014slicing}.
\end{enumerate}
 
However, the quality of the assessment depends on the methodology that is applied, which, in turn, depends heavily on data availability.

\section{Policy Implications of GVCs}
Engagement in GVCs would not bring economic prosperity in an automated manner. It requires much more value added from India’s potential productive factors and upgrading quality \& quantity of those factors with a strong distributional aspect of socioeconomic opportunities and outcomes. These challenges truly create the scope for policy discussion. However, the policy options need a strategic framework to maximise the gains from GVC engagement. To develop an effective and sustainable strategy of GVC participation, we must identify key binding constraints and suggest the necessary policy and regulatory interventions as well as infrastructure and capacity building. That's why this part of my thesis points out three distinct focus areas with corresponding objectives and challenges \citep{10}.

\subsection{Entering GVCs}
    
This first focus discusses ways for India and its peers to enter global production networks. Those avenues include ways to attract foreign investors, as well as strategies to enhance the participation of domestic firms in GVCs. Suggestions for entering GVCs encompass measures to ensure that India can offer world-class links to the global economy and create a friendly business climate for foreign tangible and intangible assets.
    
    \begin{enumerate}
        \item Creating World-Class GVC Links -- jump-starting GVC entry through the creation of EPZs and other competitive spaces \citep{31}, attracting the “right” foreign investors as foreign investors vary in their potential to deliver spillovers \citep{farole2014making}, helping domestic firms find the “right” trade partners and technology abroad, improving connectivity to international markets \citep{oecd-wto}.

        \item Creating a World-Class Climate for Firms' Assets -- ensuring cost competitiveness while avoiding the trap of low-cost tasks, improving drivers of investment and protecting foreign assets, organising domestic value chains and improving the quality of infrastructure and services \citep{28}.   
    \end{enumerate}

\subsection{Expanding and strengthening GVC participation}
    
    This second focus discusses ways for India and its comparator countries to lever their position in GVCs to achieve higher value addition through economic upgrading and densification. The concept of economic upgrading is largely about gaining competitiveness in higher-value-added products, tasks, and sectors. Densification involves engaging more local actors (firms and workers) in the GVC network. Strengthening GVC–local economy links, absorptive capacity, and skills contributes to the overall goal to increase India’s value added that results from GVC participation.

    \begin{enumerate}
        \item Strengthening GVC–Local Economy Links on the Buyer’s and Seller’s Sides \citep{farole2014making}.
        
        \item Strengthening Absorptive Capacity -- maximizing the absorption potential of local actors to benefit from GVC spillovers \citep{farole2014making}, Fostering Innovation and Building Capacity \citep{28}.
        
        \item Creating a world-class Workforce through skill development \citep{28}.
    \end{enumerate}
    
\subsection{Turning GVC participation into sustainable development}
    
    The third one focuses on social and environmental sustainability of India's GVCs. Labour market–enhancing outcomes for workers at home and more equitable distribution of opportunities and outcomes create social support for a reform agenda aimed at strengthening India’s GVC participation. Climate-smart policy prescriptions and infrastructure can mitigate the challenges for firms from climatic disruptions, ensuring the long-term predictability, reliability, and time-sensitive delivery of goods necessary to participate in GVCs.
    
    \begin{itemize}
        \item Creating a World-Class Workforce -- promoting skill development, social upgrading, and equitable distribution of opportunities and outcomes.
        \item Implementing Climate-Smart Policies and Infrastructure through proper environmental regulation and innovation in green technologies.
    \end{itemize}

\section{Summary of Statistical Analysis}

Now I have to provide a summary of the probable methodologies available to carry out the GVCs' assessment and their content in my thesis.

\begin{itemize}
    \item \underline{\textbf{Macroeconomic trends:}} Value added by broad sector, employment by broad sector, labor productivity
by sector, FDI, exports and imports (\% of GDP), exports and imports by broad economic category, and other informed classifications.

\item \underline{\textbf{Export market share growth, push, and pull
factors:}} Export market share growth; decomposition in push and pull factors using shift-share methodologies.

\item \underline{\textbf{Measuring Competitiveness in GVCs:}} Trade in main GVCs, exports of GVC products relevant to India, top five
exports, including country dimensions and follow-up analysis of
interesting patterns (such as product-specific analysis).

\item \underline{\textbf{Network analysis:}} Worldwide trade network, country trade network for sector of interest (main buyers), country trade network for sector of interest (main suppliers), and Extension to more sectors (four or five, maximum).

\item \underline{\textbf{Trade in value-added indicators:}} Domestic value added in gross exports (total growth and by sector), decomposition, foreign value added in gross exports, domestic value added in third countries' exports, sourcing and selling patterns, value added by destination, import and export upstreamness and gap, contribution of direct and indirect domestic value added and foreign value added to gross export growth.

\item \underline{\textbf{Econometric assessment:}} Impact of structural integration in GVCs (network measure) on domestic value added embodied in exports and gross exports. Probabilistic model of entry in GVCs. Impact of GVC integration (foreign value added in gross exports, domestic value added in third countries' exports) on value added and the role of national policies. Impact of GVC integration (imported input share, export share, etc.) on labor
productivity and the role of absorptive capacity. 

\item \underline{\textbf{Role of services in GVCs:}} Zoom into the services dimension of GVC analysis

\item \underline{\textbf{Product-specific case study:}} Value chain mapping and country positioning, historical/current trends, stakeholder/actor analysis, challenges and opportunities, future implications, policy implications.

\item \underline{\textbf{Policy section:}} policy suggestions based on GVC analysis, screening of policy performance indicators etc.
\end{itemize}

Indeed, I need a sound knowledge base in policy and strategic issues related to trade in general and GVCs in particular, as well as technical skills in analyzing trade and production data at the macro and firm levels. Ideally, I should have some knowledge in GVC analysis and, most important, in depth country knowledge. If I intend to do in-depth technical analysis as well, technical tools must be learned. If I consider certain methodologies essential from the outset, then bringing in specialized technical expertise to lead those components may be useful.

\end{document}